\documentclass[useAMS,usenatbib]{mn2e}
\usepackage{graphicx}
\usepackage{graphics}
\usepackage{rotating}
\usepackage{amsmath}
\usepackage{amssymb}
\usepackage{url}
\makeatletter
\def\url@leostyle{%
  \@ifundefined{selectfont}{\def\UrlFont{\sf}}{\def\UrlFont{\small\ttfamily}}}
\makeatother
\title[Analysis of \emph{Hubble} and VLT Photometry of ESO 243-49 HLX-1]{Combined Analysis of \emph{Hubble} and VLT Photometry of the Intermediate Mass Black Hole ESO 243-49 HLX-1}

\author[S. A. Farrell et al.]{S. A. Farrell$^{1}$\thanks{E-mail:
sean.farrell@sydney.edu.au}, M. Servillat$^{2}$, J. C. Gladstone$^3$, N. A. Webb$^{4,5}$, R. Soria$^6$,\newauthor T. J. Maccarone$^{7}$, K. Wiersema$^8$,  G. K. T. Hau$^9$, J. Pforr$^{10}$, P. J. Hakala$^{11}$,\newauthor C. Knigge$^{12}$, D. Barret$^{4,5}$,  C. Maraston$^{13}$  and A. K. H. Kong$^{14}$\\
$^{1}$Sydney Institute for Astronomy (SIfA), School of Physics, The University of Sydney, NSW 2006, Australia\\
$^2$Laboratoire AIM, CEA Saclay, Bat. 709, F-91191 Gif-sur-Yvette, France\\
$^3$Department of Physics, University of Alberta, Room 238 CEB, Edmonton, AB T6G 2G7, Canada\\
$^4$Institut de Recherche en Astrophysique and Plan\'{e}tologie (IRAP), Universit\'{e} de Toulouse, UPS, 9 Avenue du colonel Roche, 31028\\ Toulouse Cedex 4, France\\
$^5$CNRS, UMR5277, 31028 Toulouse, France\\
$^6$International Centre for Radio Astronomy Research, Curtin University, GPO Box U1987, Perth, WA 6845, Australia\\
$^{7}$Department of Physics, Texas Tech University, Box 41051, Lubbock, TX 79409-1051, USA\\
$^8$Department of Physics and Astronomy, University of Leicester, University Road, LE1 7RH Leicester, UK\\
$^9$European Southern Observatory, Alonso de Cordova 3107, Santiago, Chile\\
$^{10}$NOAO, 950 North Cherry Ave., Tucson, AZ 85719, USA\\
$^{11}$Finnish Centre for Astronomy with ESO (FINCA), V\"{a}is\"{a}l\"{a}ntie 20, University of Turku, FIN-21500 Piikki\"{o}, Finland\\
$^{12}$School of Physics and Astronomy, University of Southampton, Hampshire SO17 1BJ, UK\\
$^{13}$Institute of Cosmology and Gravitation, University of Portsmouth, Dennis Sciama Building, Burnaby Road, Portsmouth PO1 3FX,\\ UK\\
$^{14}$Institute of Astronomy and Department of Physics, National Tsing Hua University, Hsinchu 30013, Taiwan
}
\begin{document}


\pagerange{\pageref{firstpage}--\pageref{lastpage}} \pubyear{2002}

\maketitle

\label{firstpage}

\begin{abstract}

In this paper we present a combined analysis of data obtained with the \emph{Hubble Space Telescope} (\emph{HST}), Very Large Telescope (VLT), and \emph{Swift} X-ray telescope (XRT) of the intermediate mass black hole ESO 243-49 HLX-1 that were taken 2 months apart between September and November 2010. Previous separate analyses of these data found that they were consistent with an irradiated accretion disc with contribution from either a very young or very old stellar population, and also indicated that the optical flux of the HLX-1 counterpart could be variable. Such variability could only be attributed to a varying accretion disc, so simultaneous analysis of all data sets should break the degeneracies in the model fits. We thus simultaneously fit the broad-band spectral energy distribution (SED) from near-infrared through to X-ray wavelengths of the two epochs of data with a model consisting of an irradiated accretion disc and a stellar population. We show that this combined analysis rules out an old stellar population, finding that the SED is dominated by emission from an accretion disc with moderate reprocessing in the outer disc around an intermediate mass black hole imbedded in a young ($\sim$20 Myr) stellar cluster with a mass of $\sim$10$^5$ M$_\odot$. We also place an upper limit on the mass of an additional hidden old stellar population of $\sim$10$^6$ M$_\odot$. However, optical r'-band observations of HLX-1 obtained with the Gemini-South telescope covering part of the decay from a later X-ray outburst are consistent with constant optical flux, indicating that the observed variability between the \emph{HST} and VLT observations could be spurious caused by differences in the background subtraction applied to the two optical data sets. In this scenario the contribution of the stellar population, and thus the stellar mass of the cluster, may be higher.Nonetheless, variability of $<$ 50$\%$ cannot be ruled out by the Gemini data and thus they are still consistent within the errors with an exponential decay similar to that observed in X-rays. 

\end{abstract}


\section{Introduction}

The formation of stellar-mass black holes \citep[$\sim$3 -- 80 M$_\odot$;][]{bel10} through the collapse of massive stars is well accepted, but it is not yet completely clear how the supermassive black holes ($\sim$10$^6$ -- 10$^9$ M$_\odot$) found in the centres of galaxies are formed. Two of the most promising theories are the stellar death and the direct collapse models \citep{vol10}. Both scenarios predict that intermediate mass black holes (IMBHs) with masses between $\sim$10$^{2-5}$ M$_\odot$ will have played an important role in the formation of supermassive black holes. The existence of IMBHs also has implications for other areas of astrophysics including for the modelling of hierarchical galaxy assembly \citep[e.g.][]{hop10}, the search for dark matter annihilation signals \citep{for08}, the epoch of reionisation of the Universe \citep{ric04}, and the detection of gravitational wave radiation \citep{abb09}. The study of IMBHs and the environments in which they are found thus has important connotations for a wide range of important questions in modern astrophysics, and yet convincing evidence for their existence has until recently been scarce.

The brightest ultra-luminous X-ray source ESO 243-49 HLX-1 (hereafter referred to simply as HLX-1) currently provides the strongest evidence for the existence of IMBHs \citep{far09}. HLX-1 is located in the halo of the edge-on S0a galaxy ESO 243-49, $\sim$0.8 kpc out of the plane and $\sim$3.3 kpc away from the nucleus in projection. At the redshift of ESO 243-49 ($z$ = 0.0223) the maximum 0.2 -- 10 keV X-ray luminosity of HLX-1 is $\sim$10$^{42}$ erg s$^{-1}$ \citep{far09}, a factor of $\sim$400 above the theoretical Eddington limit for a 20 M$_\odot$ black hole. Luminosities up to $\sim$10$^{41}$ erg s$^{-1}$ can be explained by stellar-mass black holes undergoing super-Eddington accretion \citep{beg02} and/or experiencing significant beaming, which makes them appear to exceed the Eddington limit for isotropic radiation \citep{kin08,fre06}. However, luminosities above $\sim$10$^{41}$ erg s$^{-1}$ are difficult to explain without a more massive black hole. Following the discovery of an optical counterpart by \citet{sor10}, the distance to HLX-1 was confirmed through the detection of the H$\alpha$ emission line at a redshift consistent with the scenario that HLX-1 is bound to ESO 243-49 \citep{wie10}, confirming the extreme luminosity. The redshift was recently corroborated through additional spectroscopic observations by \citet{sor13b}. Independent mass estimates obtained through Eddington scaling \citep{sev11}, modelling the accretion disc emission \citep{dav11,god11b}, and the detection of ballistic jets \citep{web12} imply a mass between $\sim$9,000 -- 90,000 M$_\odot$.


Long-term monitoring with the \emph{Swift} XRT has shown that HLX-1 varies in X-ray luminosity by a factor of $\sim$50 \citep{god09}, with correlated spectral variability reminiscent of that seen in Galactic stellar-mass black holes \citep{sev11}. Since the \emph{Swift} monitoring began in 2009, HLX-1 has been observed to undergo four outbursts each spaced approximately one year apart  \citep{god11b,god12b}. The characteristic timescales of the outbursts are inconsistent with the  thermal-viscous instability model, and the outburst mechanism could instead be tidal stripping of a companion star in an eccentric orbit \citep{las11}. \citet{las11} concluded that in order to explain the implied high mass-loss rate of 10$^{-4}$ M$_\odot$ yr$^{-1}$ the donor star was most likely an asymptotic giant branch star with an initial mass of $\sim$0.5 -- 10 M$_\odot$. Higher mass stars were disregarded based on the assumption that HLX-1 resided in a globular cluster with a minimum age of $\sim$0.3 -- 0.6 Gyr. This scenario is supported by modelling of the outburst light curves under the assumption that the system is stable and the donor star will not be tidally disrupted within a few more cycles, and thus the donor star radius is not much larger than the instantaneous Roche lobe at periastron \citep{sor13}.   

Excess UV emission was detected at the position of HLX-1 with the \emph{GALEX} UV space telescope and the \emph{Swift} UV optical telescope (UVOT), although this emission could not be resolved from the nucleus of ESO 243-49 \citep{web10}. This UV excess is also consistent with the location of a pair of background galaxies at a redshift of $z$ $\sim$ 0.03, bringing into question the source of this emission \citep{wie10,far11}. In an attempt to uncover the nature of this UV excess, observations were obtained with the \emph{Hubble Space Telescope} (\emph{HST}) covering six filters from far-UV to near-IR bands following the peak of the second outburst on 13 and 23 September 2010. \citet{far12} analysed the broad-band spectral energy distribution (SED) using this \emph{HST} data in combination with simultaneous \emph{Swift} X-ray data, finding that while the X-ray spectrum was consistent with low temperature thermal emission from an irradiated accretion disc around an IMBH, the UV/optical/near-IR data were not. The addition of a component representing emission from a stellar cluster around the IMBH provided a statistically acceptable fit, but with two distinct solutions: a young ($\sim$13 Myr) stellar population with low level ($\ll$ 0.1$\%$) reprocessing in the outer disc, and an old ($\sim$13 Gyr) stellar population with extremely high ($\sim$10$\%$) reprocessing (both with stellar cluster masses of $\sim$10$^6$ M$_\odot$ and outer disc radii of $\sim$3000 times the inner disc radii). The level of reprocessing required in the old stellar population solution is borderline unphysically high, leading \citet{far12} to favour the young stellar population solution. In combination with the prominent dust lanes and lack of any nuclear activity in the host galaxy, the young stellar age and derived stellar mass suggests that HLX-1 may be the nuclear black hole remnant of a dwarf galaxy that was accreted and stripped by ESO 243-49 $<$ 200 Myr ago. 

However, Very Large Telescope (VLT) optical photometric data covering the UBVRI bands obtained on 7 and 26 November 2010 by \citet{sor12} found that the optical flux had dropped by a factor of $\sim$2. Modelling the X-ray plus optical SED they found that it was entirely consistent with a pure irradiated disc model, and was not consistent with the presence of a massive (10$^6$ M$_\odot$) young stellar population. \citet{sor12} instead suggested that if some of the optical emission were associated with a stellar cluster, it was either a massive old population ($\sim$10$^6$ M$_\odot$, 10 Gyr) or a lower mass young population of stars ($\sim$10$^4$ M$_\odot$, $<$ 6 Myr). The questions of the age and mass of the stellar cluster in which HLX-1 is embedded and therefore the origin of the IMBH is thus still open.

In this paper we report on the combined analysis of the \emph{HST} data reported in \citet{far12} and the VLT data reported in \citet{sor12} along with the associated \emph{Swift} X-ray data, fitting both data sets simultaneously with a combination of irradiated accretion disc and stellar population models. We also present an analysis of Gemini photometric monitoring data covering the third outburst of HLX-1 in an attempt to provide independent confirmation of the optical variability.

\section{Data Reduction \& Analysis}

\subsection{\emph{HST}, VLT, and \emph{Swift} Data}

For the analyses described below we use the same reduced data (obtained with the \emph{HST}, VLT, and \emph{Swift} telescopes) presented in \citet{far12} and \citet{sor12}. Both sets of \emph{Swift} spectra were binned  to a minimum of 20 counts per bin in order to use $\chi^2$ statistics for the fitting. The VLT magnitudes were converted into {\tt XSPEC}-readable spectral files using the same method outlined in \citet{far12}. For the SED fitting we utilised the {\tt XSPEC} v12.6.0q software \citep{arn96}.

We first fitted the \emph{Swift} (hereafter referred to as S2) and VLT data used in \citet{sor12} with the irradiated disc model \citep[\emph{diskir};][]{gie08,gie09}, with multiplicative components representing absorption by the neutral hydrogen column $N_H$ \citep[using the \emph{tbabs} model and the elemental abundances prescribed in][]{lod03} and dust extinction E(B-V) \citep[using the \emph{redden} model and the extinction curves in][]{car89} included. The absorption and extinction values were constrained to be greater than or equal to the Galactic values in the direction of HLX-1, i.e. 1.79 $\times$ 10$^{20}$ atoms cm$^{-2}$ \citep{kal05} and 0.013 mag \citep{sch98}, respectively. This is the same model that \citet{sor12} utilised, however they did not fit the S2 and VLT data simultaneously in {\tt XSPEC}. 

We initially froze the high-energy turn over ($kT_{e}$), Compton inner disc fraction ($f_{in}$), radius of the Compton-illuminated disc ($r_{irr}$), fraction of flux thermalised in the outer disc ($f_{out}$), and the ratio of the outer disc to inner disc radii (log($r_{\rm out}$/$r_{\rm in}$)) parameters at the same values as assumed by \citet{sor12}. The other parameters including the inner disc temperature ($kT_{disc}$), the photon index of the Compton tail ($\Gamma$), the ratio of Compton to disc luminosities ($L_c/L_d$), and the model normalisation ($N_{disc}$) were all allowed to vary freely. We obtained an acceptable fit with this model with $\chi^2_\nu$ = 0.87 for 57 degrees of freedom and model parameters all consistent within the errors with those reported in \citet{sor12}. We next thawed the $f_{out}$ and log($r_{\rm out}$/$r_{\rm in}$) values and refitted the SED, as by fitting the VLT+S1 data simultaneously we should be able to place some constraints on the properties of the outer disc. We obtained a good fit with this model with the parameters reported in Table \ref{diskirfit}. 

As reported in \citet{far12}, fitting the \emph{HST} and simultaneous \emph{Swift} (hereafter referred to as S1) data we found that the SED was not consistent with a pure disc model ($\chi^2_\nu$ = 1.6 for 30 degrees of freedom), with a clear excess detected in the far-UV that was modelled by \citet{far12} using a stellar population component. By removing the far-UV data point from our SED fitting, we can obtain a good fit with the pure irradiated disc model (see Table \ref{diskirfit} for the best fit model parameters). However, the UV excess indicates that additional components are required to account for the SED. So, although the VLT+S2 data on their own do not explicitly require a more complicated model, an additional model component representing emission from a stellar cluster is required in order to describe the $\emph{HST}$+S1 data.

\begin{table}
  \caption{Best fit parameters for the independent fitting of the \emph{HST}+S1 and VLT+S2 data with the irradiated disc model. For the \emph{HST} data, the far-UV data point was excluded. A description of each of the parameters is given in the text. The parameter values in brackets were frozen at the values indicated. The disc luminosities ($L_{0.2-10~keV}$) are over the energy range of 0.2 -- 10 keV (and are unabsorbed and de-reddened). Errors are at the 90\% confidence level.} \label{diskirfit}
  \begin{tabular}[alignment]{cccc}
  \hline  
Parameter & \emph{HST}+S1 & VLT+S2 & Units \\
  \hline
 $E(B-V)^a$ & 0.013$^{+0.05}_{-0.01}$ & 0.013$^{+0.50}_{-0.01}$ & mag \\
$N_H$ & 0.15$^{+0.07}_{-0.09}$ & 0.03$^{+0.03}_{-0.03}$ & 10$^{22}$ cm$^{-2}$ s$^{-1}$ \\
 $kT_{disc}$ & 0.17$^{+0.03}_{-0.02}$ & 0.17$^{+0.02}_{-0.05}$& keV\\
$\Gamma$ & [2.1] & 1.7$^{+0.7}_{-0.4}$\\
$kT_{e}$ & [100] & [100] & keV\\
$L_c/L_d$ & 0.07$^{+0.06}_{-0.03}$ & 0.4$^{+4.0}_{-0.3}$ & \\
$f_{in}$ & [0.0] & [0.1] & \\
$r_{irr}$ & [1.0001] & [1.2] & $r_{\rm in}$\\ 
$f_{out}^b$ & 0.0013$^{+0.0030}_{-0.0008}$ & 0.0025$^{+0.100}_{-0.002}$ & \\
log($r_{\rm out}$/$r_{\rm in}$) &  3.7$^{+0.1}_{-0.2}$ & 3.5$^{+0.5}_{-4.0}$ & \\
$N_{disc}$ & 100$^{+200}_{-100}$ & 40$^{+30}_{-10}$ & \\
$L_{0.2-10~keV}$ & 1.9 $\times$ 10$^{42}$ & 0.7 $\times$ 10$^{42}$ &  erg s$^{-1}$\\
 \hline
$\chi^2$/dof & 26.0/29 & 48.5/55 & \\
 \hline
  \multicolumn{4}{l}{$^a$The extinction pegged at the Galactic value of 0.013 mag for}\\
  \multicolumn{4}{l}{both sets of data.}\\
  \multicolumn{4}{l}{$^b$The fraction of flux thermalised in the outer disc pegged at}\\ 
  \multicolumn{4}{l}{the hard upper limit of the model for the VLT+S2 data.}
\end{tabular}
\end{table}

The emission from the accretion disc around HLX-1 is known to vary in flux considerably over time \citep{sev11}. In contrast, the emission from a stellar cluster should be stable, allowing us to use the variability of the disc to break the degeneracies in the model fitting encountered by \citet{far12}. We therefore fitted the \emph{HST}+S1 and VLT+S2 data simultaneously using the \emph{diskir} model with the addition of the same \citet{mar05} theoretical stellar population model\footnote{http://www.maraston.eu/Xspec\_models} used by \citet{far12}. For the simultaneous fitting we allowed the disc parameters to vary freely between the two sets of data, using the same constraints on the \emph{diskir} parameters as used by \citet{far12} and \citet{sor12}, but tied the stellar population parameters together. However, unlike in \citet{sor12} we did not freeze the fraction of flux from the inner disc that is reprocessed in the outer disc (i.e. the $f_{\rm out}$ parameter) for the irradiated disc model fitted to the VLT+S2 data. We first attempted tying the $f_{out}$ parameter between the two epochs of data, but found that the model tended to over-predict the disc emission in the VLT bands. This could possibly imply that the disc geometry (i.e. the disc height, solid angle etc.) and/or the albedo varied between the observations. We thus decided to allow $f_{out}$ to vary freely. We also kept the outer disc radius parameter log($r_{\rm out}$/$r_{\rm in}$) tied between the two data sets, as the luminosity of the thermal component has been seen to vary as $L_{d} \sim T_{in}^4$ \citep[][consistent with emission from the inner part of an accretion disc]{sev11} and the X-ray decline between the two epochs is exponential, which is expected once supply to the disc ceases if the  outer disc radius is constant \citep{kin98}. Absorption and extinction were accounted for as described above, with the parameters tied between the two data sets\footnote{We should note that we experimented with allowing the $N_H$ to vary freely but found that while the degrees of freedom were reduced the model fitting was not improved. As such, we chose to link the parameters between the two data sets under the assumption that the N$_H$ likely does not vary significantly over the timescale between the two sets of observations.}. 

\begin{figure*}
\begin{center}
\includegraphics[width=\textwidth]{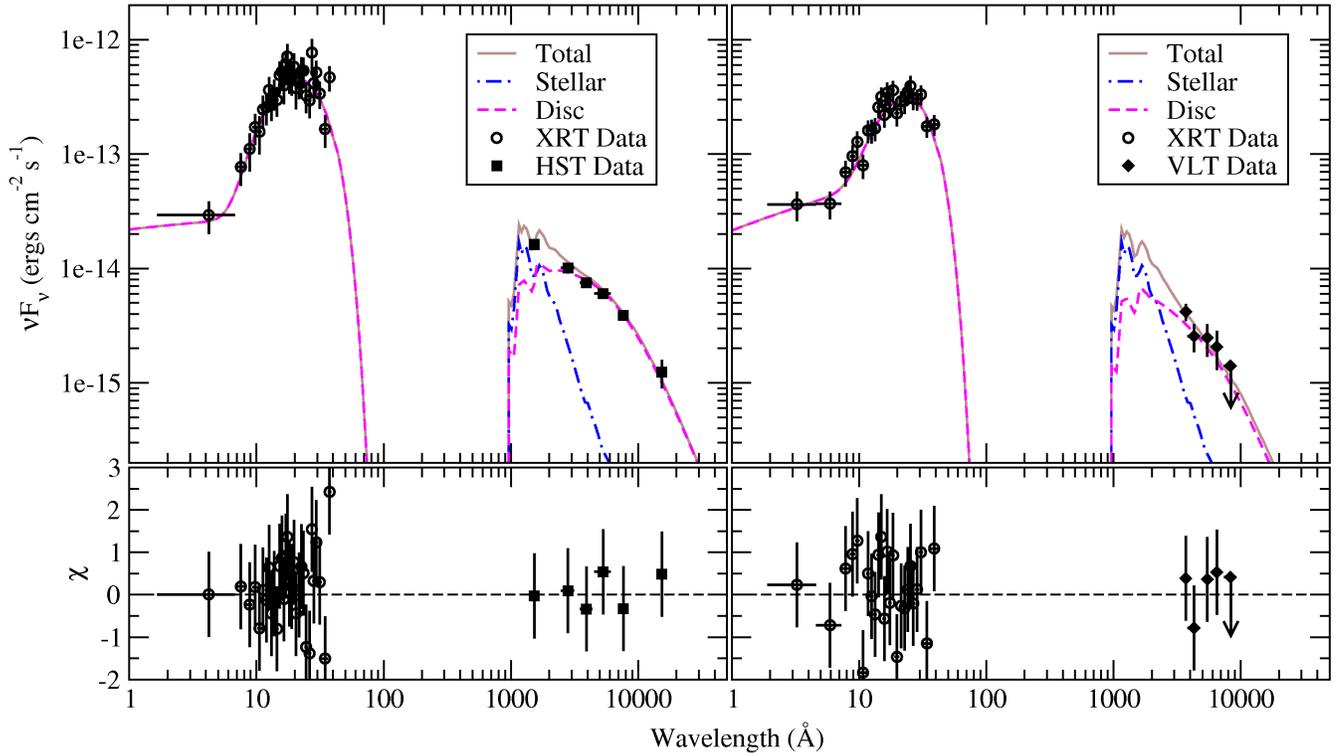}
\caption{Best-fit broadband SED model of HLX-1 constructed using the \emph{Swift} XRT, \emph{HST}, and VLT data fitted with the stellar population and irradiated disc models. \emph{Left:} the \emph{HST}+S1 data. \emph{Right:} the VLT+S2 data. The bottom panels show the fit residuals.} \label{sed1}
\end{center}
\end{figure*}

We obtained an excellent fit with the irradiated disc plus the \citet{mar05} stellar population model, with the best fit providing physically plausible parameter values (see Table \ref{specpar_mar}). The best fit SED models are shown in Figure \ref{sed1}. It should be noted that there is a large disparity between the size of the error bars in the \emph{HST} and VLT data, so the \emph{HST} data dominates the fit. In addition, the largest contribution from the stellar population model is in the UV bands, which are not sampled by the VLT data, thus introducing the far-UV $\emph{HST}$ data point has a disproportionate influence on the model fitting. As noted above, without the $\emph{HST}$ far-UV  data point both SEDs are consistent with a pure irradiated accretion disc model, highlighting the importance of UV data in order to correctly model the broad-band SED. 

\begin{table}
  \caption{Best fit parameters for the simultaneous fitting of the \emph{HST}, VLT, and \emph{Swift} data with the irradiated disc plus stellar population models, with the disc component allowed to vary freely but the stellar population fixed between the two epochs of data. A description of each of the parameters is given in the text. The parameters values in brackets were frozen at the values indicated. The luminosity for the stellar population is bolometric, while the disc luminosities ($L_{0.2-10~keV}$) are over the energy range of 0.2 -- 10 keV (all luminosities are unabsorbed and de-reddened). The luminosity and thus the mass of the stellar population component are highly unconstrained, thus we quote only the best fit value here. Errors are at the 90\% confidence level.} \label{specpar_mar}
  \begin{tabular}[alignment]{cccc}
  \hline  
Parameter & \emph{HST}+S1 & VLT+S2 & Units \\
  \hline
  \multicolumn{4}{c}{Extinction and Absorption}\\
  \hline
$E(B-V)^a$ & \multicolumn{2}{c}{0.013$^{+0.12}_{-0.00}$} & mag \\
$N_H$ & \multicolumn{2}{c}{0.04$^{+0.03}_{-0.02}$} & 10$^{22}$ cm$^{-2}$ s$^{-1}$ \\
  \hline
  \multicolumn{4}{c}{Stellar Population}\\
  \hline
$Z_*$ & \multicolumn{2}{c}{[1.0]} & Z$_\odot$ \\
$Age$ &  \multicolumn{2}{c}{2$^{+6}_{-2}$} & 10$^7$ yr \\ 
$z$ & \multicolumn{2}{c}{[0.0223]} &\\
$L_*$ & \multicolumn{2}{c}{$\sim$2 $\times$ 10$^{40}$}&  erg s$^{-1}$\\ 
$M_*$ & \multicolumn{2}{c}{$\sim$9 $\times$ 10$^{4}$} & M$_\odot$\\  
  \hline
  \multicolumn{4}{c}{Irradiated Accretion Disc}\\
  \hline
$kT_{disc}$ & 0.20$\pm$0.02 & 0.17$^{+0.01}_{-0.02}$& keV\\
$\Gamma$ & [2.1] & 1.8$^{+0.7}_{-0.5}$\\
$kT_{e}$ & [100] &[100] & keV\\
$L_c/L_d$ & 0.11$^{+0.08}_{-0.07}$ & 0.3$^{+2.0}_{-0.2}$ & \\
$f_{in}$ & [0.0] & [0.1] & \\
$r_{irr}$ & [1.0001] & [1.2] & $r_{\rm in}$\\ 
$f_{out}$ & 3.7$^{+0.3}_{-2.0}$ & 0.8$^{+0.3}_{-0.8}$ & 10$^{-3}$\\
log($r_{\rm out}$/$r_{\rm in}$) &  \multicolumn{2}{c}{4.0$^{+0.4}_{-0.2}$} & \\
$N_{disc}$ & 30$^{+20}_{-10}$ & 50$^{+50}_{-20}$ & \\
$L_{0.2-10~keV}$ & 1 $\times$ 10$^{42}$ & 0.8 $\times$ 10$^{42}$ &  erg s$^{-1}$\\
 \hline
$\chi^2$/dof &  \multicolumn{2}{c}{73.79/86} & \\
 \hline
  \multicolumn{4}{l}{$^a$The extinction pegged at the Galactic value of 0.013 mag.}
\end{tabular}
\end{table}

The metallicity ($Z_*$) of the stellar population could not be constrained and so we froze it to Solar values \citep[the metallicity of ESO 243-49 has not previously been measured, so we chose $Z_*$ = 1 $Z_\odot$ as it is the approximate midpoint of the metallicity range in the][model]{mar05}. The best fit age of the stellar population was found to be $\sim$20 Myr, with the luminosity ($L_*$) and thus stellar mass ($M_*$) lower than that found from fitting the \emph{HST}+S1 data alone. Changing the metallicity to $Z_*$ = 0.2 $Z_\odot$ or $Z_*$ = 2.0 $Z_\odot$ does not change the other parameters of the fit significantly, indicating that the fit is completely insensitive to metallicity. The stellar mass of the best combined fit (calculated using the assumed metallicity, the derived age and bolometric luminosity, and the model mass-to-light ratio\footnote{See http://www.maraston.eu}) was found to be 9 $\times$ 10$^{4}$ M$_\odot$, more than an order of magnitude lower than reported by \citet{far12}. It should be noted, however, that the large uncertainties in the luminosity (in part due to the large errors in the absorption and extinction), age, and metallicity of the stellar population produce very large errors in the stellar mass. Taking into account the errors on luminosity and age, and the allowable ranges of metallicity in the stellar population model (i.e. $Z_*$ = 0.2 -- 2.0 $Z_\odot$), we very roughly constrain the mass to be between $\sim5 \times 10^2 - 6 \times 10^6$ M$_\odot$. 

\begin{figure}
\begin{center}
\includegraphics[width=\columnwidth]{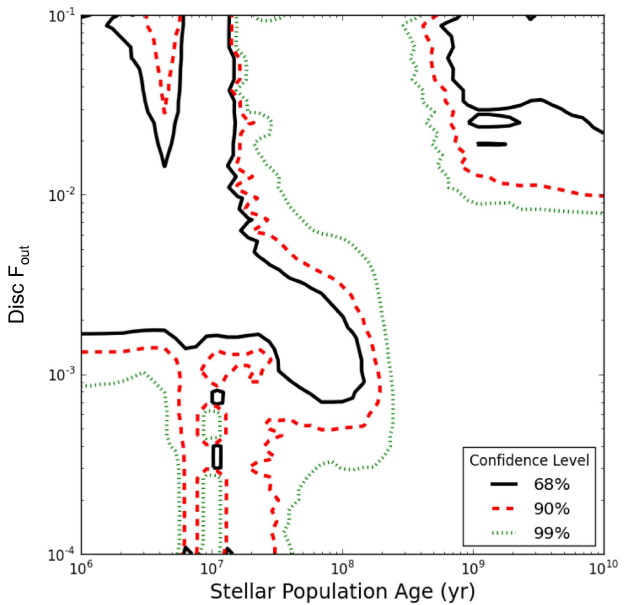}
\caption{$\chi^2$ contour plot of the stellar population age \citep[using the][theoretical stellar population model]{mar05} versus the fraction of flux thermalised in the outer disc (f$_{out}$) for the fit to the \emph{HST}+S1 data \citep[the fit parameters are those listed in][]{far12}. 
} \label{cont1}
\end{center}
\end{figure}

\begin{figure}
\begin{center}
\includegraphics[width=\columnwidth]{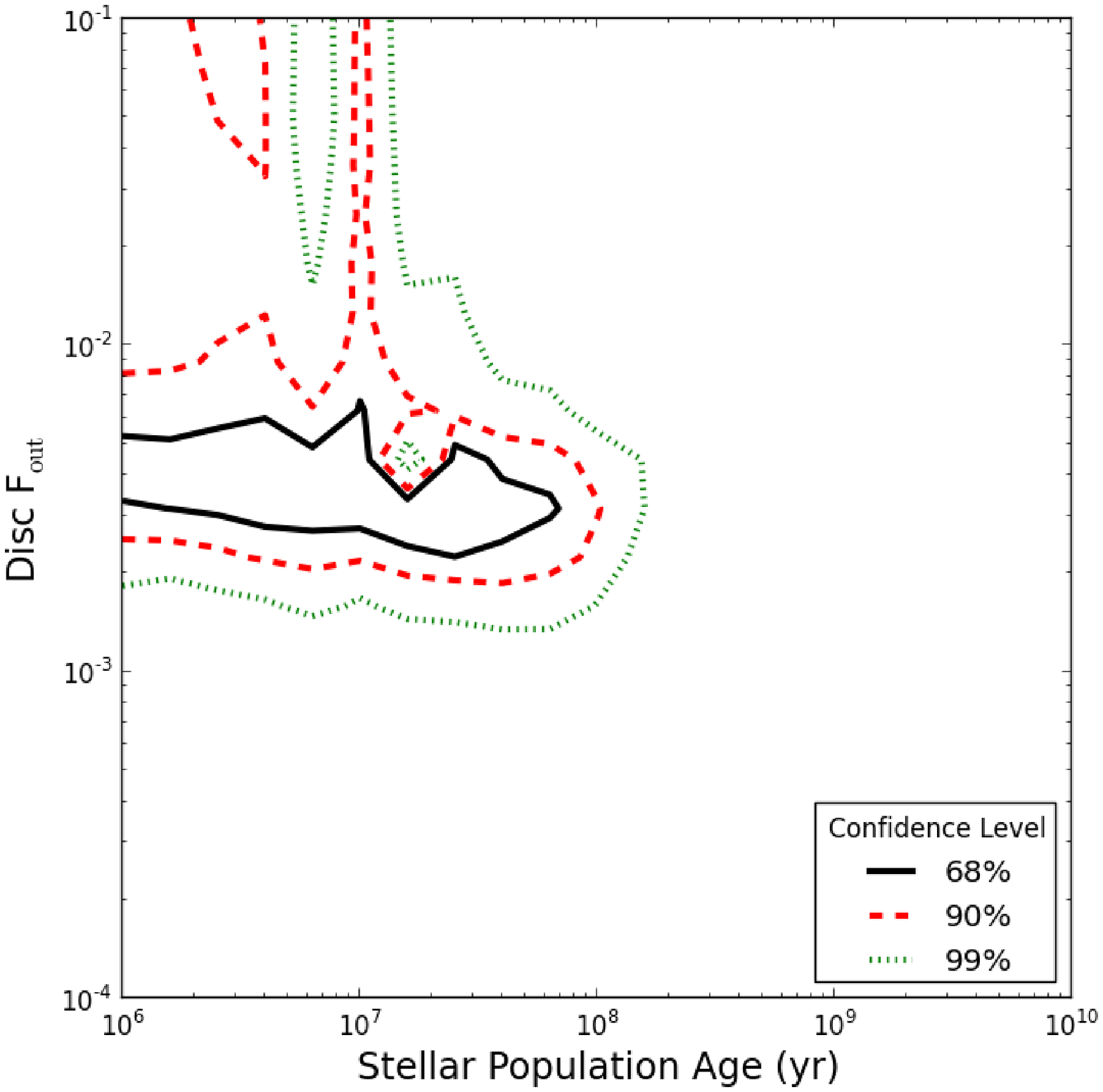}
\caption{$\chi^2$ contour plot of the stellar population age versus the fraction of flux thermalised in the outer disc (f$_{out}$) for the simultaneous fit to the \emph{HST}+S1 and VLT+S2 data. The contours shown are for the irradiated disc model fitted to the \emph{HST} data. 
} \label{cont2}
\end{center}
\end{figure}

In contrast to the results of fitting the \emph{HST}+S1 alone, by fitting the $\emph{HST}$+S1 and VLT+S2 data simultaneously we found that the old stellar population solution is no longer viable. The $\chi^2$ contour plots of the stellar population age against $f_{\rm out}$ are presented in Figure \ref{cont1} and Figure \ref{cont2}, showing that while an old ($>$ 1 Gyr) stellar population is viable when only considering the $\emph{HST}$+S1 data alone, when fitting the $\emph{HST}$+S1 and the VLT+S2 data simultaneously the old stellar population solution is no longer acceptable. The reason for this is that significant variability is observed between the red end of the $\emph{HST}$ and VLT data, which by definition requires the disc to make a significant/dominant contribution at that end of the SED, thus ruling out the possibility of an old population (which would only be able to contribute at that end of the SED). These results thus support the conclusions of \citet{far12} but with a lower stellar mass \citep[as suggested by][]{sor12}. 

The stellar population model component, however, only accounts for the dominant stellar population. Our results thus imply that there was a recent burst of star formation in the cluster $\sim$20 Myr ago, and while older stars are almost certainly also present young stars currently dominate the stellar light. Although the statistics of our data are insufficient to allow us to attempt to fit for an additional old stellar contribution, we attempted to place limits on the luminosity and thus mass of the underlying old population. Taking our best fit solution to the SED, we added an additional stellar population model with the metallicity frozen at 1.0 Z$_\odot$, the stellar age frozen at 10 Gyr, and the redshift frozen at 0.0223, and steadily increased the normalisation of this component until we could no longer obtain an acceptable fit at the 90$\%$ confidence level. In this manner we found that the upper limit on the bolometric luminosity (unabsorbed and de-redenned) of the old stellar population was $\sim$10$^{39}$ erg s$^{-1}$, giving a mass upper limit of $\sim$10$^6$ M$_\odot$. Thus, while the light from the stellar cluster appears to be dominated by a young population of stars, the data do not exclude a lower luminosity population of older stars dominating the mass of the cluster.

The irradiated disc models fitted to both the \emph{HST}+S1 and VLT+S2 data (for the best fit disc plus young stellar population solution) had physically plausible parameters broadly consistent with those reported previously by \citet{far12} and \citet{sor12}. The accretion disc temperatures agree within the errors with those reported by \citet{far12} and \citet{sor12}, as well as the $L_{disc}$ $\propto$ $T^4_{in}$ trend reported by \citet{sev11} and expected for a geometrically thin, optically thick accretion disc \citep{sha73}. The disc luminosities were also very similar to those reported previously, as were the ratios of Compton tail to disc luminosities ($L_c/L_d$). The inner disc radii were calculated from the normalisations of the irradiated disc models reported in \citet{sor12} ($N_{disc}$ = 40.6$^{+91.7}_{-13.5}$), Table \ref{specpar_mar}, and the best fit normalisation of $N_{disc}$ = 60$^{+90}_{-40}$ obtained for the young stellar population solution reported in \citet{far12} using the relationship $N_{disc}$ = [($r_{in}$/km)/($D$/10kpc)]$^2$cos$i$ and an assumed distance of $D$ = 95 Mpc. For disc angle of inclinations ranging from face-on ($i$ = 0$^\circ$) to $i$ = 75$^\circ$ (constrained via the absence of eclipses in the HLX-1 \emph{Swift} XRT light curve), the inner disc radii agree between \citet{far12}, \citet{sor12}, and this work (see Table \ref{radii}) with values around 10$^5$ km, indicative of a black hole mass of $\sim$10$^{4}$ M$_\odot$. 

In contrast, the fraction of flux thermalised in the outer disc ($f_{out}$) differed significantly from the values obtained from fitting the $\emph{HST}$+S1 data alone by \citet{far12} and the value assumed by \citet{sor12}, although still well within the physically plausible range. The ratio of the outer to inner disc radii, log($r_{out}/r_{in}$), obtained from the combined fit to the $\emph{HST}$+S1 and VLT+S2 data was also larger than the values obtained by \citet{far12} and assumed by \citet{sor12}, indicating a larger outer disc radius of $\sim$10$^{8-9}$ km. Such a large outer disc radius is impossible to reconcile with the rise and decay timescales observed in the $\emph{Swift}$ XRT light curve, which suggest outer disc radii of $\sim$10$^{7}$ km \citep[assuming a viscosity parameter of $\alpha$ $\sim$ 0.1 -- 1.0, similar to that observed in Galactic black holes][]{las11,sor12}.

However, we caution that these results are dependent upon the observed variability in the optical bands between the \emph{HST} and VLT observations being real, yet the different background subtraction methods applied to the two optical data sets could produce apparent variability. In the next section we present an analysis of optical monitoring data obtained with the Gemini-South telescope following the third X-ray outburst, in order to test whether the observed variability between the \emph{HST} and VLT observations is real (assuming that the optical flux and spectrum variations are identical from outburst to outburst).

\begin{table}
\caption{Comparison of inner and outer accretion disc size constraints between the results obtained by \citet{far12}, \citet{sor12}, and the SED fits reported in this paper.} \label{radii}
  \begin{tabular}[alignment]{lcc}
  \hline  
Data & $r_{in}$ & $r_{out}$ \\
 & (10$^{5}$ km) & (10$^{8}$ km) \\
  \hline
  \multicolumn{3}{c}{i = 0$^\circ$}\\
 \hline
$\emph{HST}$+S1 \citep{far12} & 0.4 -- 1.2 & 1.1 -- 2.9 \\
$\emph{HST}$+S1 (this work) & 0.4 -- 0.7 & 2.7 -- 16.9\\
VLT+S2 \citep{sor12} & 0.5 -- 1.1 & 1.5 -- 3.1 \\
VLT+S2 (this work) & 0.5 -- 1.0 & 3.3 -- 23.9 \\
  \hline
  \multicolumn{3}{c}{ i = 75$^\circ$}\\
 \hline
$\emph{HST}$+S1 \citep{far12} & 0.8 -- 2.3 & 2.1 -- 5.7 \\
$\emph{HST}$+S1 (this work) & 0.8 -- 1.3 & 5.3 -- 33.2\\
VLT+S2 \citep{sor12} & 1.0 -- 2.1 & 2.9 -- 6.0 \\
VLT+S2 (this work) &1.0-- 1.9 & 6.5 -- 46.9 \\

\hline
\end{tabular}
\end{table}

\subsection{Gemini GMOS observations}

We observed ESO 243-49 with GMOS-S at Gemini-South \citep[Program ID: GS-2011B-Q-31; see][for an instrument description]{hoo04} using the r' filter (r\_G0326, which covers the redshifted H$\alpha$ emission line at $\sim$6720 \AA) between 2011 August 31 and and 2011 October 6 (see Table \ref{gemobs}). The observations were performed in photometric weather conditions, in order to detect a faint point source on top of the galaxy emission, and were triggered during the month after the 2011 X-ray outburst of HLX-1. The detector pixel scale is 0.146\arcsec\ per pixel and we used a 9-point dither pattern repeated twice. 

\begin{table*}
\caption{Gemini GMOS r'-band filter observations of ESO 243-49 HLX-1.} \label{gemobs}
  \begin{tabular}[alignment]{ccccccc}
  \hline  
Night  & MJD & Date & Exp. time [s]  & Airmass & Seeing & Relative Flux\\
  \hline
1 &  55804.267 & 2011-08-31T06:25:17.1 & 4832.82 & 1.05 & 0.83 & 0.94$\pm$0.36 \\
2 & 55810.182 & 2011-09-06T04:22:20.6 & 4832.82 &  1.13 & 0.67 & 0.81$\pm$0.31 \\
3 & 55828.205 & 2011-09-24T04:55:49.4 & 4832.82 & 1.05 & 0.45 & 1.20$\pm$0.34 \\
4 & 55840.100 & 2011-10-06T02:23:37.5 & 4027.35 &  1.14 & 0.66 & 1.05$\pm$0.34\\
\hline
\end{tabular}
\end{table*}

We used the THELI pipeline\footnote{http://www.astro.uni-bonn.de/$\sim$theli/} \citep{erb05} to process the images (using bias and twilight flat images), align the images on the USNO-B1 catalogue, and stack the dithered exposures. All images were then cropped to have a similar size and pixel scale. We selected seven isolated point like stars in the field and estimated the encircled energy at different extraction radii (2 to 25 pixels). The sky was estimated in an annulus (25 to 27 pixels) and subtracted. From those stars, we also estimated the relative scale between the different exposures. The relative error for the scaling was 6\% in the rÕ band.

\begin{figure}
\begin{center}
\includegraphics[width=\columnwidth]{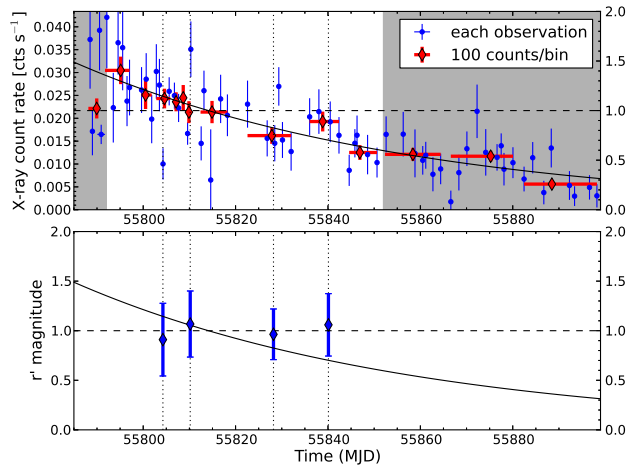}
\caption{\emph{Swift} XRT (top panel) and Gemini r'-band (bottom panel) light curves of HLX-1. The right axis in the top panel shows the normalised flux. The dashed lines indicate the mean X-ray (top) and optical (bottom) fluxes over the period not shaded in the top panel. The black lines indicate the exponential fit to the binned X-ray light curve.}
\label{gemini_lc}
\end{center}
\end{figure}

We extracted the flux from the optical counterpart to ESO 243-49 HLX-1. We subtracted the extended emission of the galaxy locally using using a thin plate spline method as implemented with the IDL procedure GRID\_TPS \citep{bar93}. This function interpolates a set of values over a regular two dimensional grid, from irregularly sampled data values, e.g. excluding a circular region of radius 5-7 pixels around the source in an image stamp. We then performed aperture photometry (radii of 4 pixels, encircling 50\% to 70\% of the total energy), and corrected for losses as estimated from the encircle energy of reference stars. The extracted fluxes were then normalised to the mean of all 4 r'-band values. The errors include the error on the flux measurement (quadratic sum of the scatter in background values, the random photon noise, and the uncertainty in mean sky brightness) and the
uncertainties from scaling and aperture corrections. The values are reported in Table \ref{gemobs} and in Figure \ref{gemini_lc} (bottom panel). The relative errors are thus about 30-40\%, due to the fact that the source is faint and buried in the galaxy emission. An absolute measurement would include additional uncertainties, which is why we chose to use data from one instrument only and focused on relative errors in this work.

It is, however, possible to get a rough estimate of the
magnitude. We used our seven reference stars and their magnitudes
reported in the GSC 2.3 catalog for the RF band
(Fmag in the catalogue). The conversion from RF to rÕ magnitudes
is not colour dependent for point sources, so we can
apply an additive correction of +0.4 mag to the Fmag values
to get rÕ-band magnitudes (see e.g. Drimmel et al. 2004).
We combined all 4 rÕ-band images and found a zero point
of 28.33 from our reference stars, with a 1 sigma error of 0.2.
This is consistent with the zeropoint given on the GeminiÊ
webpages\footnote{http://www.gemini.edu/sciops/instruments/performance-monitoring/data-products/gmos-n-and-s/photometric-zero-points} for GMOS-S (28.3--28.4). We applied this zero point to the extracted flux of the target
and combined the errors to estimate an average rÕ-band
AB magnitude of 24.12 $\pm$ 0.3 over the 4 rÕ-band images. Assuming
no colour dependence (and in fact the colour term for the
HLX-1 counterpart is lower than the errors), this converts
to a Vega R-band magnitude of 23.9 $\pm$ 0.3, consistent with the
value at the end of August 2009 \citep[23.8 $\pm$ 0.25][]{sor10} and in September 2010 \citep{far12}, but brighter than the magnitude in November 2010, three months after the outburst \citep[24.7 $\pm$ 0.4][]{sor12}.

We show in Figure \ref{gemini_lc} the X-ray count rate from our follow-up program with the $\emph{Swift}$ XRT over a similar period of time. We used the web based interface\footnote{http://www.swift.ac.uk/user\_objects/} \citep{eva09} to extract the count rate for each observation, and with a binning of 100 counts per bin. The count rates were then normalised to the mean count rate over MJD 55792 and 55852 (the area not shaded in Figure \ref{gemini_lc}, top panel). We fitted the X-ray light curve with an exponential decay, and overplot the same function over the rÕ-band light curve (Figure \ref{gemini_lc}, bottom panel).  It is clear from Figure \ref{gemini_lc} that the r'-band fluxes are consistent with both a constant and an exponential decay.

\section{Discussion \& Conclusions}

\subsection{The Combined SED Fitting}
In this paper we have presented a combined analysis of \emph{HST}, VLT, and \emph{Swift} observations of the intermediate mass black hole ESO 243-49 HLX-1 taken $\sim$2 months apart during the decay following the outburst in 2010. By fitting the VLT+S2 data simultaneously we confirm that these data are entirely consistent with an irradiated disc model without the need for an additional stellar component. However, the \emph{HST}+S1 data are not consistent with such a model, with a clear UV excess present in the SED \citep[as previously reported in][]{far12}. When fitted without the far-UV data point, the \emph{HST}+S1 SED is well described by the irradiated disc model, highlighting the importance of UV data for disentangling the disc and stellar population contributions.

In an attempt to break the degeneracies posed by these two model components we fitted the  \emph{HST}+S1 and VLT+S2 data simultaneously with a model representing emission from an irradiated disc plus a stellar population. The parameters of the disc components derived from the SED fitting are entirely physical, obtaining inner disc temperatures of 0.17 -- 0.20 keV (consistent with previous results and with emission from a geometrically thin, optically thick accretion disc), reprocessing fractions of 0.08 -- 0.37$\%$, inner disc radii of $\sim$10$^{4-5}$ km, and an outer disc radius of $\sim$10$^{8-9}$ km, all consistent with an accretion disc around a $\sim$10$^4$ M$_\odot$ black hole. The presence of reprocessing in the outer accretion disc would definitively rule out beaming as the cause of the high X-ray luminosity of HLX-1, indicating that the emission must be isotropic and thus strongly supportive of the presence of an intermediate mass black hole. 

We found that the simultaneous SED fitting of both epochs of data is inconsistent with the old stellar population solutions derived by \citet{far12} and \citet{sor12}, suggesting instead a stellar age of $\sim$20 Myr. The best fit solution indicates a higher contribution by an irradiated accretion disc than found by  \citet{far12}, resulting in a lower stellar luminosity and thus a lower mass for the stellar cluster of $\sim$10$^5$ M$_\odot$.  However, the data do not exclude the presence -- in addition to the population of young stars that dominate the stellar light -- of a lower luminosity population of older ($\sim$10 Gyr) stars with a total stellar mass up to $\sim$10$^6$ M$_\odot$. It is well known that composite populations containing a large gap in stellar ages are difficult to disentangle, because the energetic emission of stellar populations with ages smaller than $\sim$100 Myr is orders of magnitudes larger than those from older stars, and dominates the spectrum at all optical and near-IR wavelengths \citep[see e.g.][Figure~12]{mar10}. These results continue to support the scenario whereby HLX-1 is the remnant of a dwarf galaxy that has been accreted and stripped of most of its mass through a merger event with ESO 243-49.

\subsection{Constraints on the Nature of the Donor Star}
Having a young stellar population opens up the possibility of having a supergiant or Wolf-Rayet donor star. In both cases, the high mass loss rates could be sufficient to power HLX-1. Additionally, there is good evidence, both theoretical \citep[see][]{fra02} and observational \citep[see e.g.][]{smi02} that the outer accretion disc radii in systems fed by fast winds may be smaller than those fed by Roche lobe overflow. In the event that the system is wind-fed, the possibility then remains that the fast rise timescales of the outbursts might be explained without requiring the extremely large eccentricities and the small radii invoked in \citet{sor13}. This scenario will be investigated further in a future paper (Miller et al., in preparation).

\subsection{The Optical Variability}

The apparent change in optical brightness between different epochs strongly implies that the irradiated disc provides the dominant contribution to the optical continuum (not including the far-UV band). Specifically, $R \approx 23.8 \pm 0.1$ mag in the \emph{HST} observations (obtained 
by interpolating the optical continuum between the V and I bands) and 
$R \approx 23.9 \pm 0.3$ mag in the Gemini data (both observed about a month after the X-ray outburst), but $R \approx 24.7 \pm 0.4$ mag in the VLT images (three months after outburst).
However, it is possible that this optical variability may be artificial, 
a result of the different background subtraction methods applied. 
For the \emph{HST} data, the background light from ESO 243-49 was simply subtracted from a nearby
region using standard aperture photometry, as the superior angular
resolution of the \emph{HST} allows us to resolve out the background
contribution from the host galaxy. The VLT data of ESO 243-49 were modelled
by fitting the local galaxy background around HLX-1 with a thin plate spline
model (similar to the approach we adopted for the Gemini GMOS data, with
similar magnitudes derived) and then subtracted prior to measuring the
photometry. The errors were bootstrapped from this fitting procedure and
thus include the effects of uncertainty in the local background. This is
crucial, since the photometric errors are dominated by the local background
estimate and any background estimation error would introduce spurious
variability when comparing to the \emph{HST} data.

Gemini observations in the r'-band covering part of the decay of the third
X-ray outburst in 2011 show no evidence for any significant variability in
optical bands but do not rule out variability $<$ 50$\%$.  Nonetheless, an
exponential decay similar to that observed in X-rays is not ruled out by the
Gemini data. We also compared each r'-band observation with the closest
X-ray observation in time (1h, 2h, 5h and 7h difference for nights 1 to 4,
respectively). No correlation was observed, though we note that this could
be due to the low number of counts detected in each \emph{Swift} XRT
observation. In addition, fluctuations in the binned X-ray light curve are
not correlated with the r'-band light curve.

We stress that the Gemini data do not rule out the presence of variability,
but instead indicate that it is important to compare optical data taken with
the same telescope and instrument setup and using the same background
subtraction method in order to confirm the variability and thus constrain
the parameters of the SED model components. Further observations with the
\emph{HST} or longer term monitoring with ground based 8m class telescopes
at different X-ray luminosities are thus necessary in order to definitively
break the degeneracies in the SED fitting and thus confirm the age and mass
of the stellar cluster, as well as the fraction of reprocessing in the
accretion disc.

\section*{Acknowledgments}
We thank the anonymous referee for their comments that improved this paper as well as Tal Alexander, Cole Miller, and Chris Done for useful discussions. SAF is the recipient of an Australian Research Council Postdoctoral Fellowship, funded by grant DP110102889. MS acknowledges support from the Centre National dÕEtudes Spatiales (CNES) and from program number HST-GO-12256, which was provided by NASA through a grant from the Space Telescope Science Institute, which is operated by the Association of Universities for Research in Astronomy, Inc., under NASA contract NAS5-26555. JCG acknowledges support from the Avadh Bhatia Fellowship and Alberta Ingenuity. Based on observations made with the NASA/ESA \emph{HST} associated with program 12256. Based on observations made with ESO Telescopes at the Paranal Observatory under program ID 088.D-0974(B). Based on observations obtained at the Gemini Observatory, which is operated by the Association of Universities for Research in Astronomy, Inc., under a cooperative agreement with the NSF on behalf of the Gemini partnership: the National Science Foundation (United States), the National Research Council (Canada), CONICYT (Chile), the Australian Research Council (Australia), Minist\'{e}rio da Ci\^{e}ncia, Tecnologia e Inova\c{c}\~{a}o (Brazil) and Ministerio de Ciencia, Tecnolog\'{i}a e Innovaci\'{o}n Productiva (Argentina). We thank the \emph{Swift} team for granting us Target of Opportunity observations in support of this program.


\bsp

\label{lastpage}

\end{document}